\documentstyle[preprint,aps]{revtex}

\begin{document}

\draft
\title
{Fractal Statistics and Quantum Black Hole Entropy}

\author
{Wellington da Cruz\footnote{E-mail: wdacruz@exatas.uel.br}}

\address
{Departamento de F\'{\i}sica,\\
 Universidade Estadual de Londrina, Caixa Postal 6001,\\
Cep 86051-970 Londrina, PR, Brazil\\}
 
\date{\today}

\maketitle

\begin{abstract}

Simple considerations about the fractal characteristic of 
the quantum-mechanical path give us the opportunity to derive 
the quantum black hole entropy in connection with the concept 
of fractal statistics. We show the geometrical origin of the 
numerical factor of four of the quantum black hole entropy expression and the 
statistics weight appears as a counting of the quanta of geometry.

\end{abstract}

\pacs{PACS numbers: 05.30.-d; 04.70.Dy; 04.70.Bw; 04.60.-m\\
Keywords: Fractal statistics; Quantum black hole}

The fractal properties of path swept by 
point-particles are described by the relation $L\sim R^h$, i.e. between 
a closed path of length $L$ and size $R$, with $h$ a fractal 
parameter or Hausdorff dimension. On the other hand, the values 
of $h$ can be obtained from the propagators of the particles\cite{R1}. 
We have introduced the concept of universal classes of 
particles or quasiparticles labeled just by this fractal parameter, 
such that for $h$ defined in the interval 
$1$$\;$$ < $$\;$$h$$\;$$ <$$\;$$ 2$, we obtain {\it Fractal Statistics} 
( fractal distribution function ) 
given by\cite{R2}

\begin{eqnarray}
\label{e.h} 
n=\frac{1}{{\cal{Y}}[\xi]-h},
\end{eqnarray}

\noindent where the function ${\cal{Y}}[\xi]$ satisfies 
the equation 

\begin{eqnarray}
\xi=\left\{{\cal{Y}}[\xi]-1\right\}^{h-1}
\left\{{\cal{Y}}[\xi]-2\right\}^{2-h},
\end{eqnarray}
 
\noindent and $\xi=\exp\left\{(\epsilon-\mu)/KT\right\}$ 
has the usual definition. Thus we generalize {\it in a natural way} 
the fermionic($h=1$) and bosonic($h=2$) distributions for 
particles carrying rational or irrational values for 
the spin quantum number $s$. 
The particles termed {\it fractons} ( they live in two-dimensional 
multiply connected space ) are collected 
in each class $h$ taking into account the {\it fractal spectrum}

\begin{eqnarray}
&&h-1=1-\nu,\;\;\;\; 0 < \nu < 1;\;\;\;\;\;\;\;\;
 h-1=\nu-1,\;
\;\;\;\;\;\; 1 <\nu < 2;\\
&&etc.\nonumber
\end{eqnarray}

\noindent and the spin-statistics relation $\nu=2s$.

We emphasize that the Fractal Statistics captures the observation about the 
{\it fractal nature} of the quantum-mechanical path, which reflects the 
Heisenberg uncertainty principle. Besides this, the classes $h$ satisfy a 
{\it duality symmetry} defined by ${\tilde{h}}=3-h$ and so, we extract a 
{\it fractal supersymmetry} which defines pairs of particles $\left(s,s+
\frac{1}{2}\right)$. In this way, our formulation can be understood as a {\it 
quantum-geometrical} description of the statistical laws of Nature.

In this Letter, we show that simple considerations about the fractal 
characteristic of the quantum path give us the quantum black 
hole entropy expression. Following Hausdorff, we 
define a length $L$\footnote{ The 
Hausdorff length is scale independent ( self-similar ) and has 
fractal dimension\cite{R3}.}  as 

\begin{eqnarray}
\label{e.1}
L=l_{p}R^h,
\end{eqnarray}

\noindent where $l_{p}=\sqrt{\frac{\hbar G}{c^3}}$ 
is the Planck length. Squaring the eq.(\ref{e.1}) and taking the 
logarithm, we obtain

\begin{eqnarray}
\ln{R}=\frac{1}{2h}\ln{\frac{A}{l_{p}^2}},
\end{eqnarray}

\noindent  with $A=L^2$(area).

Given that the quantum of gravitational field is bosonic, $h=2$ and 
as $\frac{A}{l_{p}^2}$ is a sufficiently large number, 
we assume that $\frac{A}{l_{p}^2}=N!$ ( $A\gg 10^{-66}$ and $N$ 
is considered as a large number ) and using 
the Stirling$^\prime$s formula

\begin{eqnarray}
\ln{R}&=&\frac{1}{4}\ln{N!}\nonumber\\
&=&\frac{N}{4}\ln\left[\frac{N}{e}\right]\nonumber\\
&=&\frac{A}{4l_{p}^2(N-1)!}\ln\left[\frac{N}{e}\right],
\end{eqnarray}

\noindent and  

\begin{eqnarray}
\frac{\ln{R}^{N!}}{\ln{N!}}
\equiv \ln{\cal{W}}={\frac{A}{4l_{p}^2}}
\equiv\frac{\cal{S}}{K},
\end{eqnarray}

\noindent ($K$ is the Boltzmann constant), hence ${\cal{W}}=
R^{\frac{N!}{\ln N!}}$, with $N!=R^{4}=\frac{A}{l_{p}^2}$, i.e.\\ 
${\cal{W}}=\left\{N!\right\}^{\frac{N!}{4\ln N!}}=
\left\{\frac{A}{l_{p}^2}\right\}^{\frac{A}
{4l_{p}^2\ln\left\{\frac{A}{l_{p}^2}\right\}}}=
e^{\frac{A}{4l_{p}^2}}=e^{\frac{\cal{S}}{K}}$, \\

is the 
statistics weight of the macroscopic state of entropy $\cal{S}$ 
or the total number 
of microscopic states ({\it quanta of geometry} ) compatible 
with this macroscopic state. Thus, we can say that $\cal{W}$ 
establishes a notion of 
{\it quantum geometry}. Therefore, in this way we identify 
the quantum black hole 
entropy as ( {\it quantum} here means that the Hausdorff dimension $h$ 
encodes information about the quantum path and 
the particle associated to it )\cite{R4}\footnote{ Another direction of research, 
exploiting an analogy between edge states in $(2+1)$-dimensional 
gravity and fractional quantum Hall efffect, also has extracted as 
result a geometric entropy\cite{R5}.}

\begin{eqnarray}
{\cal{S}}_{BH}&=&K\ln{\cal{W}}\\
&=&\frac{Kc^3}{4\hbar G}A,
\end{eqnarray}

\noindent where the quantities $c,G,\hbar,$ are the speed of light, 
Newton$^\prime$s constant, Planck$^\prime$s constant and we note that 
the numerical factor of four has a {\it geometrical origin}
\footnote{ A discussion about this 
coefficient by Bekenstein, properly, reinforces our topological 
interpretation of it\cite{R6}.}. 
Finally, in some sense the {\it holographic principle}\footnote{ 
The entropy in the region $A$ is bounded by its boundary, i.e. the 
closed length $L$ or the dynamics of a system in $A$ can be 
described by a system which lives in $L$.}\cite{R7} 
is realized in our approach, since the fractal parameter 
$h$ is related to 
the boundary($L\sim R^{h=2}$) of the area($A\sim R^{2h=4}$), 
so in one phrase {\it all 
in ${\cal{S}}_{BH}$ is pure geometry}\footnote{ It is possible, 
classically, that the phase space of a system gravitating a black hole be 
compact, with volume scaling as area ( external to black hole ) 
and so, the holographic 
principle is a consequence of the gravitating system 
quantization\cite{R8}.}. It is worth noting that our analysis 
is supported by the results of another consideration about $2$-point 
function in two-dimensional quantum gravity
\cite{R9}.

\end{document}